\documentclass[usenatbib,onecolumn]{mn2e}
\usepackage{newtxtext,newtxmath}
\usepackage{lipsum}
\usepackage[T1]{fontenc}
\usepackage{ae,aecompl}
\usepackage{hyperref}
\usepackage{graphicx}	
\usepackage{amsmath}	
\usepackage{amssymb}	
\usepackage{bm}	
\usepackage{caption}
\usepackage{multirow}
\usepackage{multicol}
\usepackage{cuted}
\usepackage{mathtools}
\usepackage{longtable}

\title[Probabilistic galactic dynamics I]{Probabilistic galactic dynamics I -- the Sun and GJ 710 with Monte Carlo, linearised and unscented treatments}
\author[F. Feng]
{F. Feng$^{1}$\thanks{E-mail: f.feng@herts.ac.uk or fengfabo@gmail.com}, H. R. A. Jones$^{1}$\\
$^{1}$Centre for Astrophysics Research, University of Hertfordshire, College
Lane, AL10 9AB, Hatfield, UK}
\date{\today}

\begin{document}
\maketitle 
\begin{abstract}
Deterministic galactic dynamics is impossible due to the space-time randomness caused by gravitational waves. Instead of treating stellar orbits deterministically, we integrate not only the mean but also the covariance of a stellar orbit in the Galaxy. As a test case we study the probabilistic dynamics of the Sun and the star GJ 710 which is expected to cross the Oort Cloud in 1.3\,Myr. We find that the uncertainty in the galactic model and the Sun's initial conditions are important for understanding such stellar close encounters. Our study indicates significant uncertainty in the solar motion within 1\,Gyr and casts doubt on claims of a strict periodic orbit. In order to make such calculations more practical we investigate the utility of the linearised and unscented transformations as two efficient schemes relative to a baseline of Monte Carlo calculations. We find that the linearised transformation predicts the uncertainty propagation as precisely as the Monte Carlo method for a few million years at least 700 times faster. Around an order of magnitude slower, the unscented transformation provides relative uncertainty propagation to a very high precision for tens of millions of years. There exist a variety of problems in galactic dynamics which require the propagation of the orbital uncertainty for more than one or two objects and the first order linearised transformation provides an efficient method which works to Gyr time scales for small initial uncertainty problems and for propagation over hundreds of million years for larger initial uncertainty problems. 
  \end{abstract}
\section{Introduction}     \label{sec:introduction}
In the nineteenth century, Pierre-Simon Laplace claimed that ``nothing would be uncertain and the future just like the past would be present before its (Laplace's demon) eyes''\citep{laplace12}. However, such a strong determinism assumes the reversibility of a system which seems to be incompatible with the indeterminacy of quantum mechanics \citep{hermann92} and the second law of thermodynamics \citep{marcondes88}. Nevertheless, it is typically agreed that celestial dynamics is deterministic if we know the initial conditions and the system perfectly, although a system may be chaotic and unpredictable due to a lack of precise knowledge of the system. However, it is now established that space time is perturbed by a stochastic gravitational wave background, part of which is from the quantum fluctuation during the inflation period and thus is intrinsically or irreducibly random (e.g., \citealt{starobinskii79,rubakov82,abbott16}). This primordial gravitational wave background has significantly influenced the evolution of the universe \citep{grishchuk90} and stellar dynamics \citep{crosta17,moore17}. It is expected to be observed in the cosmic microwave background (CMB; e.g., \citealt{turner93,boyle08}) or be directly detected \citep{Ungarelli05,crowder05}. Therefore a probabilistic approach to celestial dynamics is unavoidable and better reflects the reality than a deterministic one.

The Monte Carlo (MC) method is conventionally employed to account for a lack of knowledge of the precise initial condition and the environment of an astrophysical system. In the case of galactic dynamics, the current position and velocity of a star are not exactly known. The acceleration experienced by a star is not well known either due to our poor knowledge of the whole Galaxy. To know the past and future orbit of a star probabilistically, one needs to draw a large number of samples from the covariance of the initial conditions and model parameters, and then integrate each orbit independently. For a system of $n$ dimensions, the number of integrations is $\mathcal{O}(N^{n})$ if N samples are drawn from each dimension. Hence the MC method becomes very expensive in computing resources for high-dimensional problems such as the probabilistic integration of planetary systems or the dynamics of stellar streams accounting for model uncertainties. As a result the MC method becomes impractical for the study of galactic dynamics in the era of large surveys such as Gaia \citep{gaiaDR2}. Although a distribution function can be used to model the Galaxy \citep{binney15}, it assumes the dynamical equilibrium of a galactic component which seems to be problematic based on recent analysis of the Gaia data \citep{antoja18}. Overall then more efficient methods are required for galactic dynamics computations in order to achieve a robust understanding of the various components and systems of interest in our Galaxy. We examine the practical implementation of treating the orbit of a star not as a single trajectory but as the propagation of probability density, similar to the evolution of the wave function in quantum mechanics.

In this work, we study the probabilistic dynamics of individual stars in the Galaxy modeled in a probabilistic way. Although in this initial work we do not account for the time variation of the galactic potential caused by substructures such as spiral arms or space-time perturbations from gravitational waves. We introduce two potentially efficient methods called the ``linearised transformation'' (LT) and the ``unscented transformation''(UT) to integrate not only the mean orbit but also the covariance of the orbit. These two methods have been frequently used to implement the Kalman filter in nonlinear systems for tracking, navigation, and robotic applications (e.g., \citealt{smith62,schmidt66,julier04}). The LT is long established though the UT method has become steadily more popular and has replaced the LT in many nonlinear filtering and control applications. We compare the LT, UT and MC methods for our initial investigation into probabilistic galactic dynamics. The numerical experiments done in this work provide a guide for the application of LT and UT in various astrophysical problems. 

We adopt the galactic model suggested by \cite{price-whelan17}, which includes a Hernquist nucleus and bulge, an Miyamoto-Nagai disk \citep{miyamoto75}, and a NFW halo \citep{nfw96}. We adopt the model parameters from \cite{price-whelan17} for the nucleus, bulge, and disk and use the halo parameters determined by \cite{watkins18} from the kinematics of globular clusters measured by Gaia and HST. Since the bulge, disk, and halo parameters are determined with about 10\%, 20\%, and 50\% uncertainty from pre-DR2 data \cite{bland-hawthorn16, mcmillan17}, we adopt these 10\%, 10\%, 20\%, and 50\% uncertainties of the mean parameter values as the uncertainties of model parameters. Since the spread in some parameters is too high or too low to be realistic if drawn from a Gaussian distribution, we truncate some of these Gaussian distributions using lower and upper bounds. We also vary the Sun's initial conditions according to the mean values and uncertainties suggested by \citep{bland-hawthorn16} based on a review of various measurements. We show the adopted parameters, uncertainties and truncations for the galactic potential model and the Sun's initial conditions in Table \ref{tab:model}.
 
\begin{table}
  \caption{Summary of the adopted properties and parameter bounds for the galactic potential and Solar characteristics. For properties with a "-" we adopt a Gaussian distribution, otherwise we show the upper and lower bounds for a truncated Gaussian distribution.}
  \label{tab:model}
  \begin{tabular}{l*7{r}}
    \hline
   Parameter &Symbol& Unit&$\bar{\bm \theta}$&$\sigma(\bm \theta)/\bar{\bm \theta}$&Lower bound & Upper bound\\\hline\hline
    Nucleus mass  &$M_{\rm n}$&$10^{10}M_\odot$&0.17&0.10&0.86&0.26\\
    Nucleus scale length  &$l_{\rm n}$&kpc&0.070&0.10&0.035&0.11\\
    Bulge mass&$l_{\rm b}$&$10^{10}M_\odot$&0.50&0.10&0.25&0.75\\
    Bulge scale length  &$M_{\rm b}$&kpc&1.0&0.10&0.50&1.5\\
Disk mass &$M_{\rm d}$&$10^{10}M_\odot$&6.8&0.20&3.4&10\\
    Disk scale length &$l_{\rm d}$&kpc&3.0&0.20&0&6\\
    Disk scale height &$h_{\rm d}$&kpc&0.28&0.20&0&0.56\\
Virial mass&$M_{\rm v}$&$10^{10}M_\odot$&167&0.50&0&418\\ 
Concentration&$M_{\rm v}$&$10^{10}M_\odot$&16.8&0.50&8.0&20\\ 
Sun's distance from the galactic Center&$R_\odot$&kpc&8.2&0.10&--&--\\
Sun's distance from the galactic Plane&$z_\odot$&kpc&0.025&0.0050&--&--\\
U component of the solar motion w.r.t. LSR&$U_\odot$&km/s&10&1.0&--&--\\
V component of the solar motion w.r.t. LSR&$V_\odot$&km/s&11&2.0&--&--\\
W component of the solar motion w.r.t. LSR &$W_\odot$&km/s&7.0&0.50&--&--\\\hline
  \end{tabular}
\end{table}

The paper is structured as follows. We introduce the LT method in section \ref{sec:lt} followed by an introduction of the UT method in section \ref{sec:ut}. In section \ref{sec:example}, we investigate the behaviour of the Sun and GJ 710 as instructive examples to compare the

we use two examples to compare the UT and LT methods and discuss the limits of various methods and draw together conclusions in section \ref{sec:conclusion}. 

\section{Linearised transformation}\label{sec:lt}
It is routine to determine the state of a linear system based on the linear propagation of mean and covariance. For a quasi-linear system, the system can be linearised locally at a given time in order to propagate uncertainty. Since the orbital period of the Sun around the Galaxy is about 200\,Myr \citep{binney11}, the acceleration of stellar motion is small over a few million years, we can approximate the stellar motion as linear over such time scales.

For this initial implementation of probabilistic galactic dynamics, we define the ``absolute uncertainty'' as the orbital uncertainty in the Galactic reference frame, and ``relative uncertainty'' in the heliocentric frame. Although relative uncertainty can be derived from the absolute uncertainty, the concept of relative uncertainty is useful since it can be generalized to assess any two objects moving relative to one another.  Considering that many stars are measured in the heliocentric frame, the orbital uncertainty of a star with respect to the Sun is more relevant in most cases and so is more appropriate than the absolute orbital uncertainty. For example, stellar encounters of the Sun are identified by finding the closest distance between a star and the Sun. Likewise, micro-lensing events are predicted based on the astrometric data of stars measured in a heliocentric frame. Thus the prediction precision is limited by the relative motions of stars rather than by absolute motions. 

Generally, the mean $\bar{\bm x}$ and covariance ${\bm \Sigma}$ of the state of a nonlinear system at time step $k$ is
\begin{eqnarray}
  \bar{\bm x}_{k}&=&{\bm f}({\bar{\bm x}_{k-1}})~,\\
  {\bm \Sigma}_{k}&=&{\bm A_{k-1}}{{\bm \Sigma}_{k-1}}{\bm A^{T}_{k-1}}~,
  \end{eqnarray}
where $A_{k-1}$ is the Jacobian of the function $\bm f$ at $\bar{\bm x}_{k-1}$. For the case of probabilistic galactic dynamics, we derive the mean and covariance of the state vector of a star $({\bm r_k}, {\bm v_k})$ from the previous step to the second order of the time step $\delta t$, 
\begin{align}
    (\bar{\bm r}_k, \bar{\bm v}_k)&= {\bm f}(\bar{\bm r}_{k-1}, \bar{\bm v}_{k-1})\nonumber\\
  {\rm cov}({\bm r_k},{\bm r_k})&= {\rm cov}({\bm r_{k-1}},{\bm r_{k-1}})+\left[{\rm cov}({\bm r_{k-1}},{\bm v_{k-1}})+{\rm cov}({\bm v_{k-1}}, {\bm r_{k-1}})\right]\delta t+{\rm cov}({\bm v_{k-1}},{\bm v_{k-1}})\delta t^2+\frac{1}{2}\left[{\rm cov}({\bm a_{k-1}},{\bm r_{k-1}})+{\rm cov}({\bm r_{k-1}},{\bm a_{k-1}})\right]\delta t^2\nonumber~,\\
  {\rm cov}({\bm r_k},{\bm v_k})&= {\rm cov}({\bm r_{k-1}},{\bm v_{k-1}})+{\rm cov}({\bm v_{k-1}},{\bm v_{k-1}})\delta t+{\rm cov}({\bm r_{k-1}},{\bm a_{k-1}})\delta t +{\rm cov}({\bm v_{k-1}},{\bm a_{k-1}}) \delta t^2+\frac{1}{2}{\rm cov}({\bm a_{k-1}},{\bm v_{k-1}}) \delta t^2\nonumber~,\\
  {\rm cov}({\bm v_k},{\bm v_k})&= {\rm cov}({\bm v_{k-1}},{\bm v_{k-1}})+\left[{\rm cov}({\bm v_{k-1}},{\bm a_{k-1}})+ {\rm cov}({\bm a_{k-1}}, {\bm v_{k-1}})\right]\delta t +{\rm cov}({\bm a_{k-1}},{\bm a_{k-1}})\delta t^2~,\label{eqn:linear}\\
  {\rm cov}({\bm r_k},{\bm \theta})&= {\rm cov}({\bm r_{k-1}},{\bm \theta})+ {\rm cov}({\bm v_{k-1}},{\bm \theta})\delta t +\frac{1}{2}{\rm cov}({\bm a_{k-1}},{\bm \theta})\delta t^2\nonumber~,\\
  {\rm cov}({\bm v_k},{\bm \theta})&= {\rm cov}({\bm v_{k-1}},{\bm \theta})+ {\rm cov}({\bm a_{k-1}},{\bm \theta})\delta t\nonumber~,
  \end{align}
where $\bm f$ is the Newtonian differential function, $\bm \theta$ are the parameters in the galactic model. The acceleration ${\bm a}_{k-1}$ is expanded to the first term in the Taylor series of the gradient of the Galactic potential $\Phi(\bm r_{k-1}, \bm \theta)$. Defining ${\bm \eta}_{k-1}\equiv(\bm r_{k-1}, \bm \theta)$, the acceleration becomes
\begin{equation}
  {\bm a}_{k-1}=-\nabla\Phi(\bm \eta_{k-1})\approx -\nabla\Phi(\bar{\bm \eta}_{k-1})+{\bm J_{k-1}}({\bm \eta}_{k-1}-\bar{\bm \eta}_{k-1})~. 
  \label{eqn:a}
\end{equation}
In the above equation, $\bm J_{k-1}$ is the Jacobian matrix of ${\bm a}_{k-1}$, which is
\begin{equation}
    {\bm J_{k-1}}= D_{\bm \eta}{\bm a}_{k-1}\bigg|_{\bm \eta=\bar{\bm \eta}_{k-1}}
  \label{eqn:jacobian}
\end{equation}
Replacing the acceleration-related covariances in Eqn. \ref{eqn:linear} with Eqn. \ref{eqn:a}, the covariances with $\bm a_{k-1}$ become
  \begin{align}
    {\rm cov}({\bm a_{k-1}},{\bm r_{k-1}})&={\bm J_{k-1}}{\rm cov}({\bm \eta_{k-1}},{\bm r_{k-1}})~,\nonumber\\
    {\rm cov}({\bm a_{k-1}},{\bm v_{k-1}})&={\bm J_{k-1}}{\rm cov}({\bm \eta_{k-1}},{\bm v_{k-1}})~.
\label{eqn:lt1}
  \end{align}
  Notably, ${\rm cov}({\bm r_{k-1}},{\bm a_{k-1}})$ and ${\rm cov}({\bm v_{k-1}},{\bm a_{k-1}})$ are the transposes of ${\rm cov}({\bm a_{k-1}}, {\bm r_{k-1}})$ and ${\rm cov}({\bm a_{k-1}},{\bm v_{k-1}})$, respectively.

  Since the acceleration is expanded using the first order Taylor series, this method is called ``first order linearised transformation'' (or LT1). For the zero order linearised transformation (LT0), the acceleration uncertainty is ignored and thus the acceleration-related covariances are equal to zero. The mean orbit estimated by LT is the orbit integrated from the initial mean state since the mean of the first order Taylor expansion of a state is zero.
  
We derive the covariance of heliocentric state $\bm h\equiv(\bm r_h, \bm v_h)$ from the covariances of the galactocentric stellar state $\bm g\equiv(\bm r_g, \bm v_g)$ and the solar state $\bm s\equiv(\bm r_\odot, \bm v_\odot)$ using
  \begin{eqnarray}
   {\rm cov}({\bm h},{\bm h})&=& {\rm cov}({\bm g},{\bm g})+{\rm cov}({\bm s},{\bm s})-{\rm cov}({\bm g},{\bm s})-{\rm cov}({\bm s},{\bm g})~,\nonumber \\
   {\rm cov}({\bm h},{\bm \theta})&=& {\rm cov}({\bm g},{\bm \theta})-{\rm cov}({\bm s},{\bm \theta})~.
  \end{eqnarray}
  The transformation from heliocentric covariance to galactocentric covariance is the reverse of the above transformation. For stars in the solar neighborhood, the relative error within a few million years is determined by the observational error which could be very small in the Gaia era. Nevertheless, the precision of relative  uncertainty propagation is limited by the first order Taylor expansion of the acceleration and thus limited by the uncertainty of galactic potential which is typically much larger than the observational error, as we will see in section \ref{sec:sun} and \ref{sec:gl710}. To avoid the direct influence of acceleration on the spatial uncertainty, we approximate ${\rm cov}({\bm r_k},{\bm r_k})$ to the first order of $\delta t$ as
  \begin{equation}
      {\rm cov}({\bm r_k},{\bm r_k})= {\rm cov}({\bm r_{k-1}},{\bm r_{k-1}})+\left[{\rm cov}({\bm r_{k-1}},{\bm v_{k-1}})+{\rm cov}({\bm v_{k-1}}, {\bm r_{k-1}})\right]\delta t~.
    \end{equation}
Thus the acceleration uncertainty can be propagated to the spatial uncertainty by changing the velocity-related covariance and the impact of the first order expansion of acceleration minimized. This modified LT1 method will be compared with the LT0 method in section \ref{sec:example}. transformations.

\section{Unscented transformation}\label{sec:ut}
The LT is the {\it de facto} method used in the Extended Kalman Filter (EKF) to determine states and infer parameters for nonlinear systems such as robotic motion planning and control (e.g. \citealt{chen12}) where computational speed and empirical results are under increasing scrutiny. However, as reviewed by \cite{julier04}, the LT or EKF is ``difficult to implement, difficult to tune and only reliable for systems that are almost linear on the time scale of the updates''. Since the mean state of LT is not updated (see Eqn. \ref{eqn:linear}), a wrong initial estimate of the system state would bias the uncertainty propagation \citep{huang10}. To overcome these limitations, the unscented transformation (UT) is used to propagate the mean state and covariance of nonlinear systems \citep{julier97}.

The UT is based on the fact that the covariance ${\rm cov}({\bm x},~{\bm x})$ can be decomposed into a set of sigma points $\bar{\bm x}$ which are derived from the square root of the covariance matrix. The nonlinear function is applied to each sigma point to determine a set of transformed points which are used to update the covariance \citep{julier97}. In the case of stellar dynamics, the UT is implemented in the following steps.
    \begin{enumerate}
    \item Estimate the initial mean state of a star $\bar{\bm x}_0=(\bar{\bm r}_0,~\bar{\bm v}_0,~\bar{\bm \theta}_0)$ (with $n$ variables in total) and the corresponding covariance ${\rm cov}({\bm x_0},~{\bm x_0})$ from the observables (e.g. proper motion, parallax, radial velocity, etc.) using the Monte Carlo method\footnote{The transformation of initial state and covariance can also be calculated using the UT. But we use the Monte Carlo estimation for various uncertainty propagation methods in order to compare them properly.}.
    \item Generate $2N+1$ weighted sigma points from the covariance matrix at the $(k-1)^{\rm th}$ time step according to
      \begin{align}
        \mathcal{X}_0 &= \bar{\bm x}_{k-1}~,\nonumber\\
           \mathcal{X}_i &= \bar{\bm x}_{k-1}+\eta{\rm cov}({\bm x}_{k-1},~{\bm x}_{k-1})_i~,\\
           \mathcal{X}_{i+n}&= \bar{\bm x}_{k-1}-\eta{\rm cov}({\bm x}_{k-1},~{\bm x}_{k-1})_i~,\nonumber
      \end{align}
where $\kappa=n(\alpha^2-1)$ and $\eta=\sqrt{n+\kappa}$ are scaling factors. According to \cite{wan00}, $\alpha\in[10^{-4},1]$ and $\beta=2$ is optimal for Gaussian distributions. We also set $\alpha=10^{-3}$ following the suggestion in \cite{wan00}. 
    \item Transform the sigma points through the Newtonian differential equations $\bm f$,
      \begin{equation}
        \mathcal{Y}_i={\bm f}(\mathcal{X}_i)~.
        \end{equation}
    \item Calculate the mean and covariance at the $k^{\rm th}$ time step,
      \begin{align}
        \bar{\bm x}_k&=\sum_{i=0}^{i=2n}W_i^{(m)}\mathcal{Y}_i\\
{\rm cov}({\bm x_k},~{\bm x_k})&=\sum_{i=0}^{i=2n}W_i^{(c)}(\mathcal{Y}_i-\bar{\bm x}_k)(\mathcal{Y}_i-\bar{\bm x}_k)^{\rm T}~,
      \end{align}
      where $\{W_i\}$ is a set of scalar weights, $W_0^{(m)}=\kappa/(n+\kappa)$, $W_0^{(c)}=\kappa/(n+\kappa) +(1+\alpha^2+\beta)$, and $W_i^{(m)}=W_i^{(c)}=1/[2(n+\kappa)]$ for $i=1,...,2n$.
      \item Repeat (ii)-(iv) for each time step. 
      \end{enumerate}
      The UT approximates the covariance and mean at least to the second order Taylor expansion \citep{wan00}. Thus the UT typically outperforms the LT although the UT is closely related to some second order LT methods \citep{gustafsson12}. We will compare these methods for probabilistic galactic dynamics in section \ref{sec:example}. 

\section{Probabilistic motion of the Sun and its close encounter GJ710}\label{sec:example}
In this section, we consider the approach of probabilistic galactic dynamics and the utility of different numerical methods for problems of immediate anthropic interest. In particular, we consider the uncertainty in the Sun’s orbit around the Milky Way as well as and the heliocentric motion of GJ 710 which is chosen because it is the star thought to produce the closest encounter with the Sun.

\subsection{Uncertainty in the Solar orbit}\label{sec:sun}
\begin{figure}
  \centering
  \includegraphics[scale=0.6]{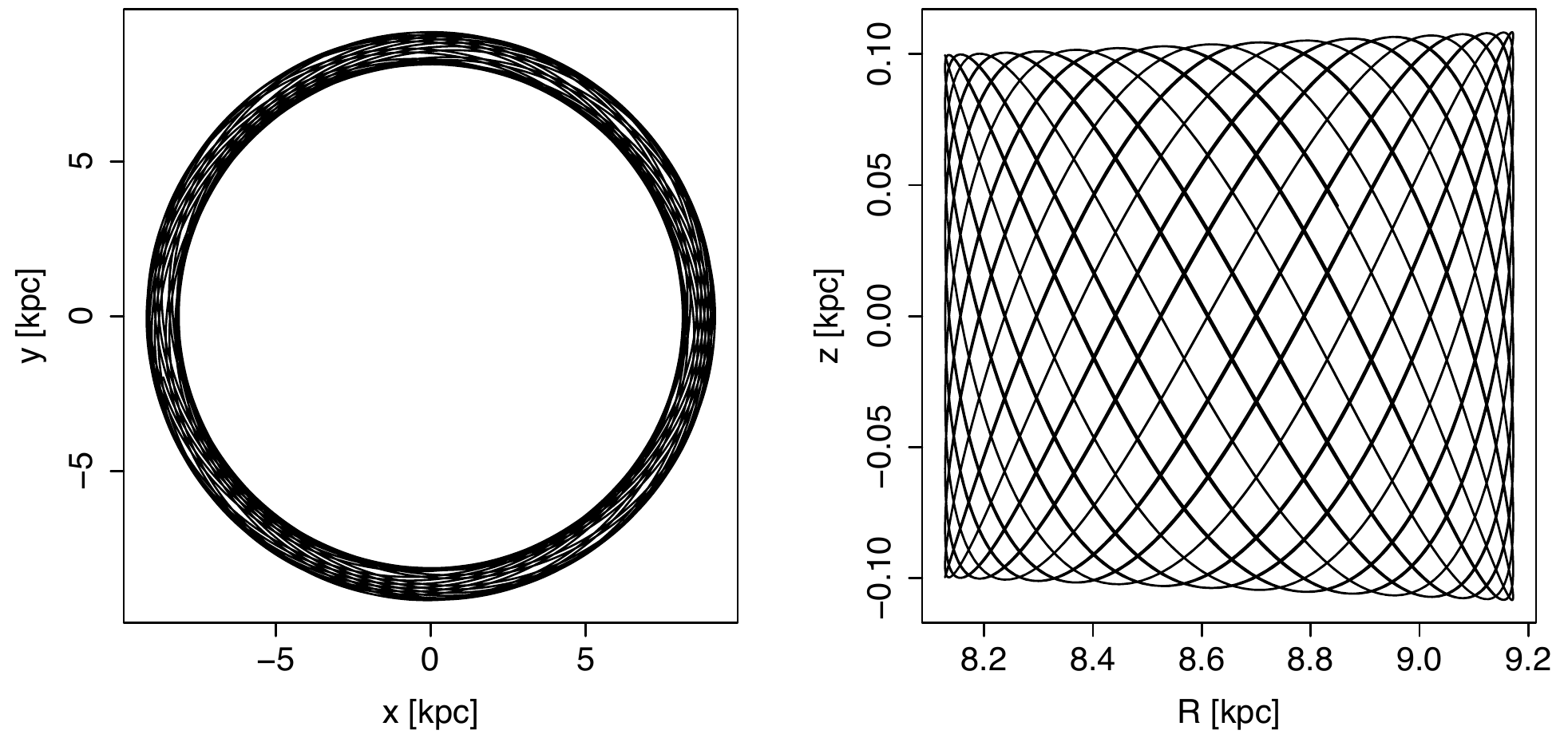}
  \caption{Example Solar orbit in the coordinates of galactic plane (left) and in the cylindrical coordinates (right).}
  \label{fig:example}
\end{figure}

To study the influence of the uncertainty in the Sun's initial conditions and the galactic potential on the solar motion, we integrate the Sun's orbit in three Monte Carlo experiments: only vary the Sun's initial condition (``SunMC''), only vary the galactic potential (``GalaxyMC''), and vary both (``SunMC+GalaxyMC''). For each experiment, we draw 1000 sets of parameters from Gaussian distributions of the parameters in Table \ref{tab:model}. We use the MC covariance of the initial conditions as the initial covariance for testing the other uncertainty propagation methods under scrutiny. Thus we ensure that that MC uncertainty propagation is the reference point for the comparison of the different uncertainty propagation methods. For each set of parameters, we integrate the orbit of the Sun forward using the Bulirsch-Stoer integrator with a time step of 1\,Myr. With the asymmetric galactic potential in section \ref{sec:introduction}, the vertical angular momentum of the Sun is conserved to a relative precision of $10^{-14}$. An example of the Solar orbit clones in $R$ and $z$ is shown in Fig. \ref{fig:example}. We see that the Solar orbit clone is confined on a torus determined by the action-angle variables. In subsections \ref{sec:5Gyr} and \ref{sec:absolute}, we will use the MC method to explore the uncertainty of the solar motion over 5\,Gyr and compare the UT though we choose to use the Monte Carlo estimation for comparison of the various uncertainty propagation methods.

\subsubsection{Probabilistic Solar orbit over 5\,Gyr}{\label{sec:5Gyr}}
To study the uncertainty propagation for the solar motion since the Sun's birth, we integrate the Sun's orbit using the MC method over 5\,Gyr. We generate 1000 clones for each of the SunMC, GalaxyMC, and SunMC+GalaxyMC experiments. The distributions of the vertical angular momentum $L_z$, the epicycle $T_R$ and vertical $T_z$ oscillation periods of the solar motion, and the Sun's orbital period $T_\phi$ are shown in Fig. \ref{fig:epicycle}. It is evident that the galactic model is the main source of orbital
uncertainty over timescales of a few Gyr. The values of $T_R$, $T_z$, and $T_\phi$ estimated based on a SunMC+GalaxyMC experiment are 163.2$\pm$16.7, 87.8$\pm$10.6, and 224.2$\pm$22.3\,Myr, respectively. The vertical oscillation period leads to a mid-plane crossing period of 43.9$\pm$5.3\,Myr, consistent with the value determined using the model including galactic spiral arms \citep{feng13}. However, this period is significantly lower than the 62\,Myr periodicity found by \cite{rohde05} in the biodiversity data and if there is periodicity in mass extinction events is inconsistent with a simple causative connection with the Solar vertical motion.
\begin{figure}
  \centering
  \includegraphics[scale=0.6]{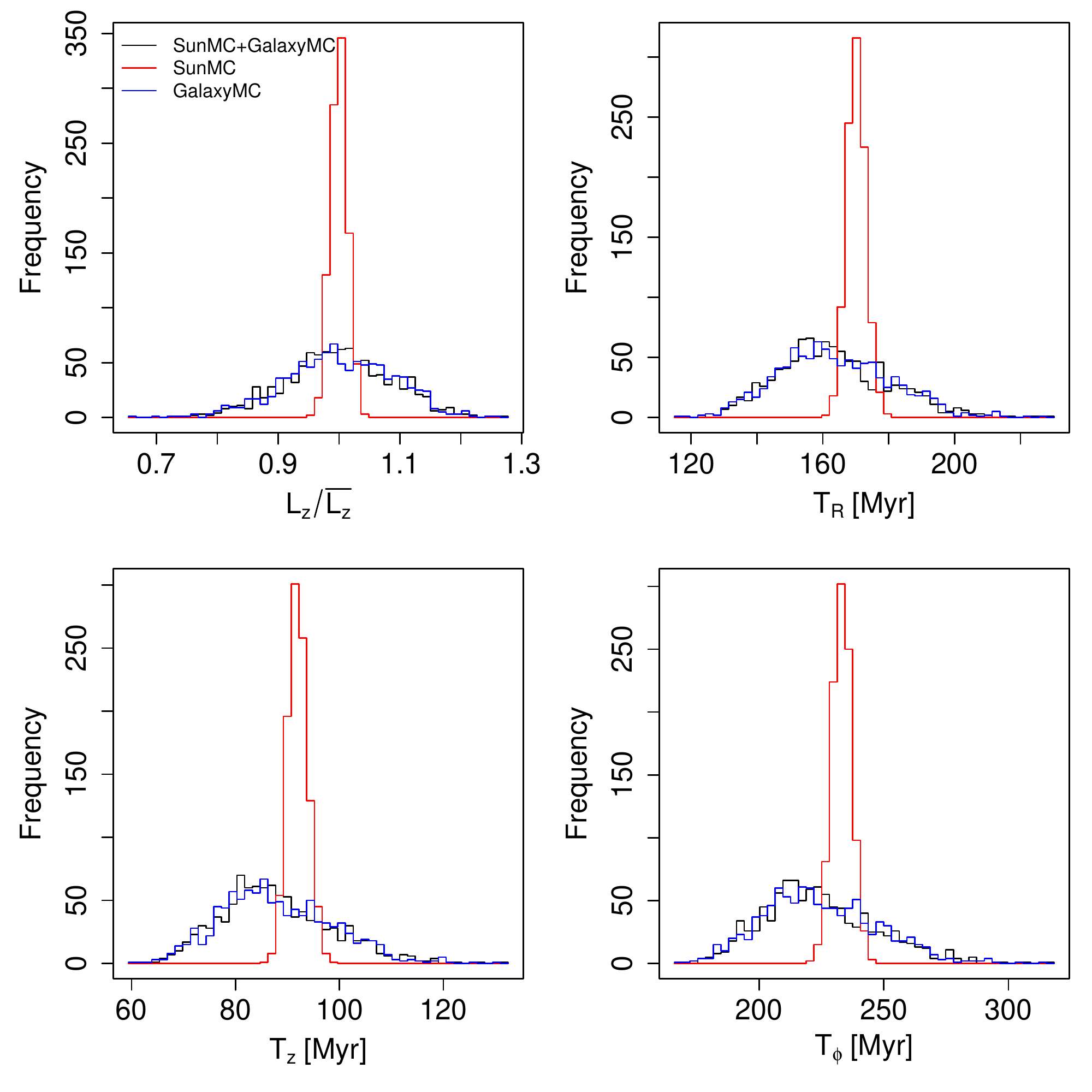}
  \caption{Distributions of $L_z/\bar{L}_z$ (top left), $T_R$ (top right), $T_z$ (bottom left) and $T_\phi$ (bottom right) of the 1000 Solar orbit clones for the three Monte Carlo experiments. The black, red and blues lines shown the distributions for the SunMC+GalaxyMC, SunMC, and GalaxyMC experiments, respectively. The parameter range is divided into 20 bins to compute the histograms. Note that the vertical angular momentum is scaled to have a unit mean. }
  \label{fig:epicycle}
\end{figure}

\begin{figure}
  \centering
  \includegraphics[scale=0.6]{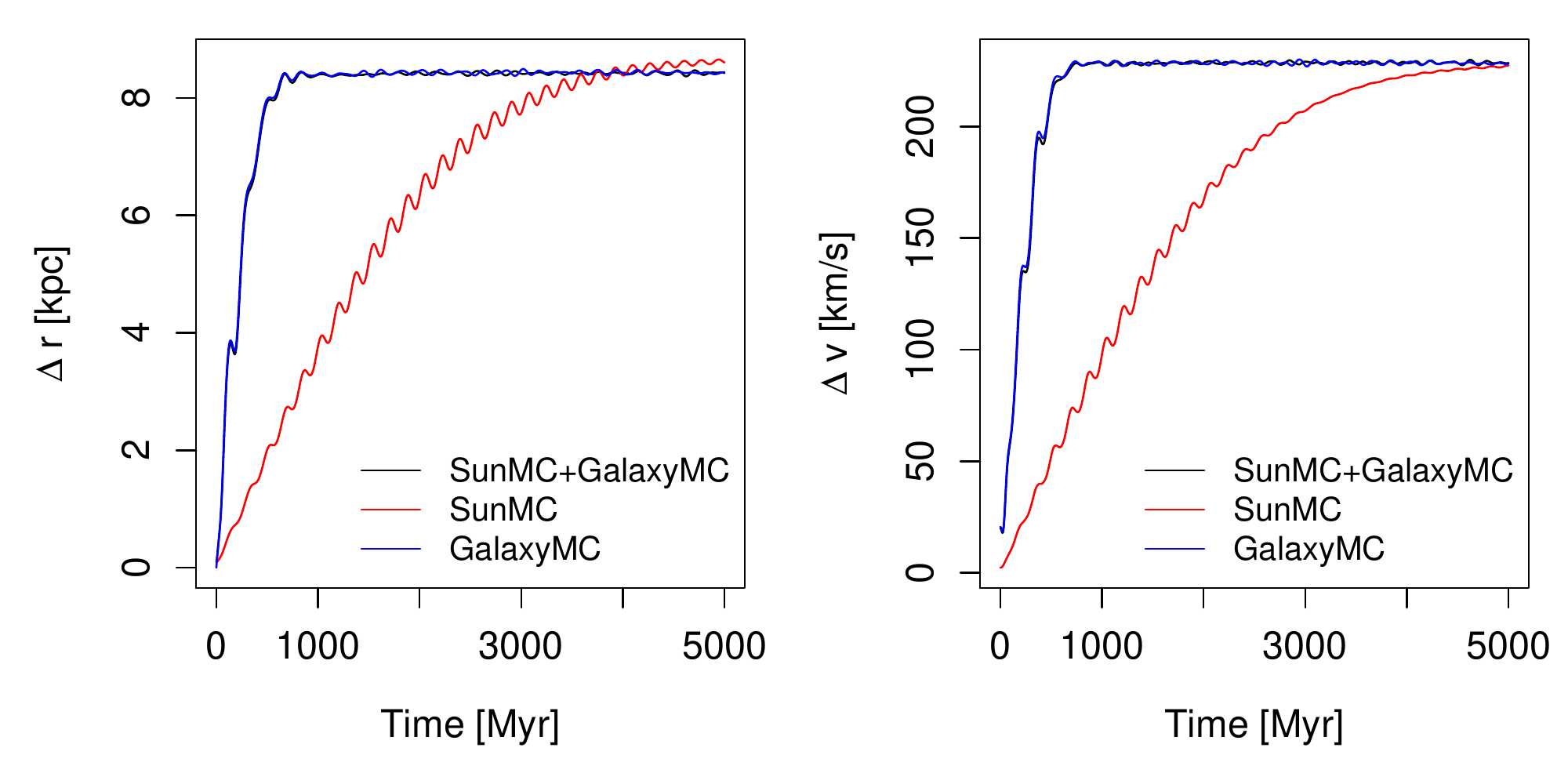}
  \caption{Spatial (left) and velocity (right) uncertainty of the Solar orbit as functions of time for the SunMC+GalaxyMC, SunMC, and GalaxyMC experiments. Assuming a Poisson distribution of Solar orbit clones, the error of $\Delta r$ and $\Delta v$ are approximately $\Delta r/\sqrt{1000}$ and $\Delta v/\sqrt{1000}$, respectively. Note that the SunMC+GalaxyMC (black) and GalaxyMC (blue) distributions are very similar and thus their difference is barely visible in the figures. }
  \label{fig:dr_dv}
\end{figure}

The uncertainty of the Solar orbit as a function of time is shown in the left panel of Fig. \ref{fig:dr_dv}. We find that the Solar orbit becomes totally unpredictable within 800\,Myr for the GalaxyMC and SunMC+GalaxyMC experiments but is partially predictable within 5\,Gyr for the SunMC experiment. Over Gyr timescales, the galactic potential is the main uncertainty source for orbital integrations. We characterise a predictability time scale $\tau$ as $\Delta r(\tau)=r_\odot(t=0~{\rm Myr})/2$ or $\Delta v(\tau)=v_\odot(t=0~{\rm Myr})/2$ of about 200\,Myr, if accounting for the uncertainty in the galactic potential. Over Gyr timescales, the galactic potential is the main uncertainty source for orbital integrations. In other words, the Solar orbit is only predictable for one Solar orbital period given our current knowledge of the Galaxy. The predictability time scale will become smaller still when other uncertainties are accounted for, e.g., galactic sub-structures such as spiral arms \citep{feng13,martinez16}. 

\begin{figure}
  \centering
  \includegraphics[scale=0.6]{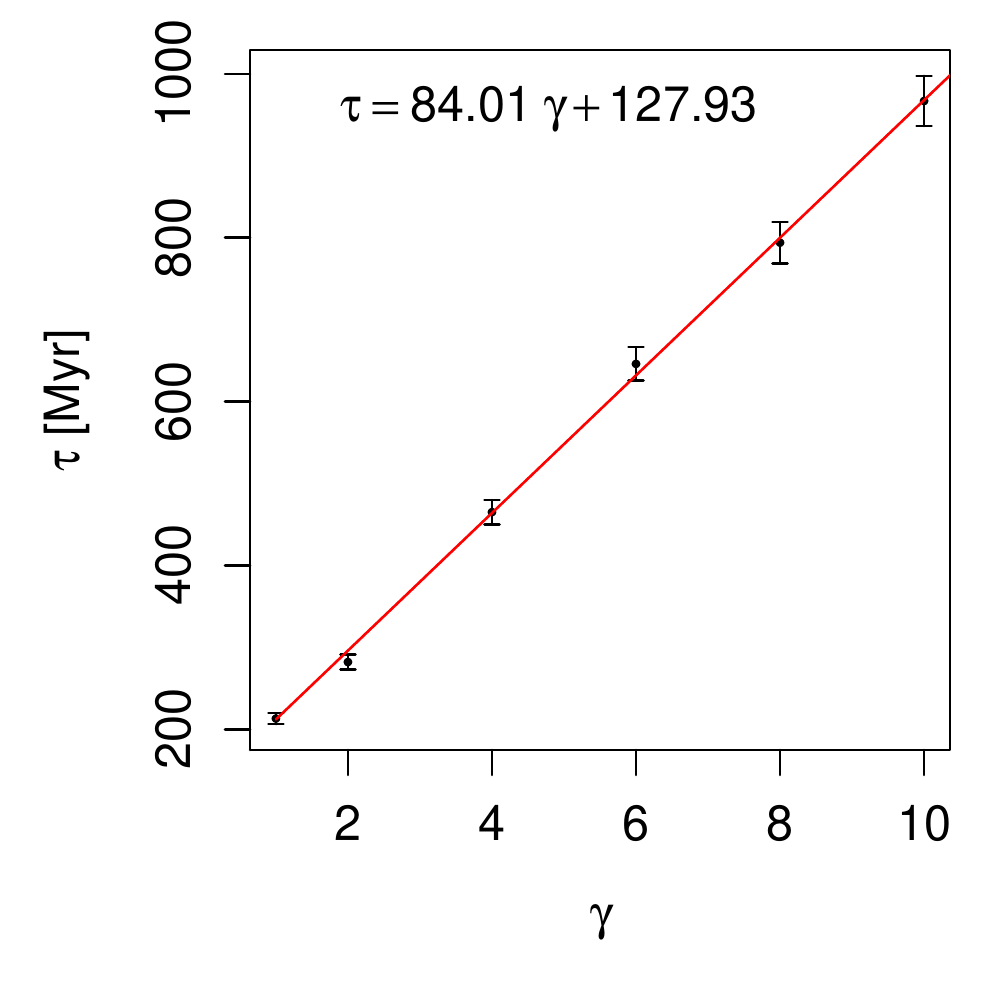}
  \caption{Predictability time scale $\tau$ as a linear function of the improvement factor of model precision $\gamma$. The error bars are calculated assuming Poisson noise in the 1000 clones of Solar orbit.}
  \label{fig:gamma}
\end{figure}

We also investigate the dependence of $\Delta r$ on model parameters by calculating their Pearson correlation coefficients with respect to $\Delta r(t=5~{\rm Gyr})$. We find that the orbital uncertainty strongly depends on the disk mass $M_{\rm d}$, disk length scale $l_d$, viral mass $M_{\rm viral}$, the concentration of NFW profile $c$, and the Solar tangential velocity $v_{y\odot}$, which have correlation coefficients of 0.53, -0.20, 0.33, 0.42, and -0.77, respectively. Thus the improvement of these parameters would significantly improve the precision of orbital integration for the Sun and the Solar neighborhood. To quantify this improvement, we reduce the uncertainty of these five parameters by a factor of $\gamma=$2, 4, 6, 8, 10, and calculate the predictability time scale $\tau$ for each. The results and the best-fit linear function are shown in Fig. \ref{fig:gamma}. We see that an increase of model precision by a factor of two would increase the predictability time scale by 84\,Myr. Since the orbital uncertainty is linearly proportional to the integration time within the predictability time scale as seen in Fig. \ref{fig:dr_dv}, the precision of the Solar orbit is linearly proportional to the model precision. In other words, to double the precision of orbital integration for the Sun requires doubling the precision of the parameters, especially of the disk and halo model as well as the tangential Solar velocity. 

Although the Solar orbit is not predictable over timescales of Gyrs, the spatial distribution of the clones of Solar orbit is probably predictable due to the conservation of angular momentum in an axisymmetric galactic potential \citep{simon09,martinez16}. We investigate this by showing the distribution of Solar orbit clones in the galactic plane at the end of the simulations (i.e. $t=5$\,Gyr). We see the clones are located within a torus because an individual orbit is conserved in $L_z$ and is described by the action-angle coordinates which define a torus \citep{binney11}. If we treat the Solar orbit clones as the orbits of Solar siblings, the regions with $7.6<R<9.0$\,kpc and $-0.07<R<0.07$\,kpc are the preferred place for finding Solar siblings. However, we expect that transient spiral arms and other sub-structures will result in larger uncertainties \citep{martinez16}. 
\begin{figure}
  \centering
  \includegraphics[scale=0.6]{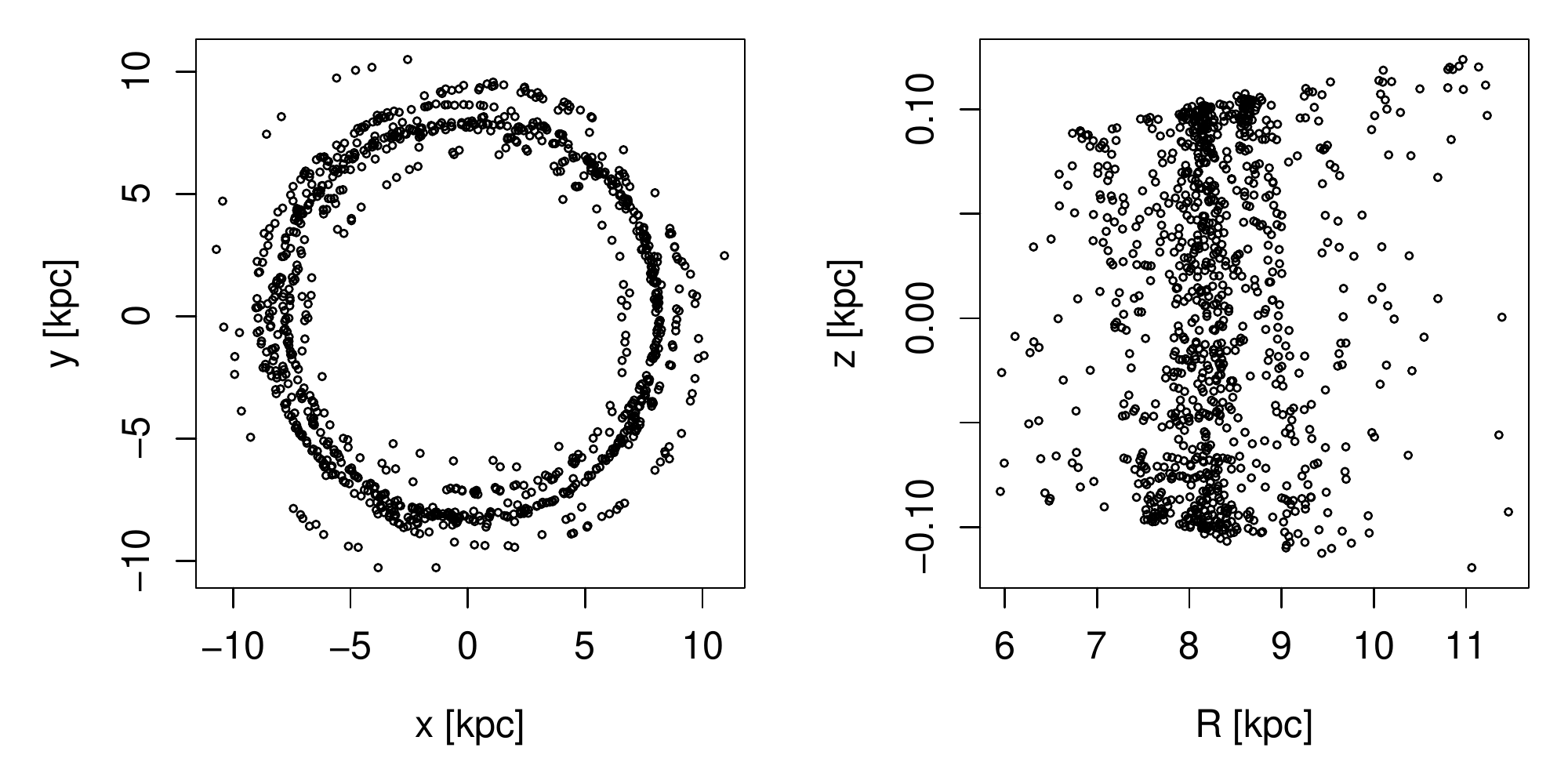}
  \caption{Distribution of the clones of Solar orbit in the galactic plane (left) and in the plane perpendicular to the galactic plane (right). }
  \label{fig:epicycle}
\end{figure}

\subsubsection{Comparison of uncertainty propagation for different methods}\label{sec:absolute}
\begin{figure}
  \centering
  \includegraphics[scale=0.40]{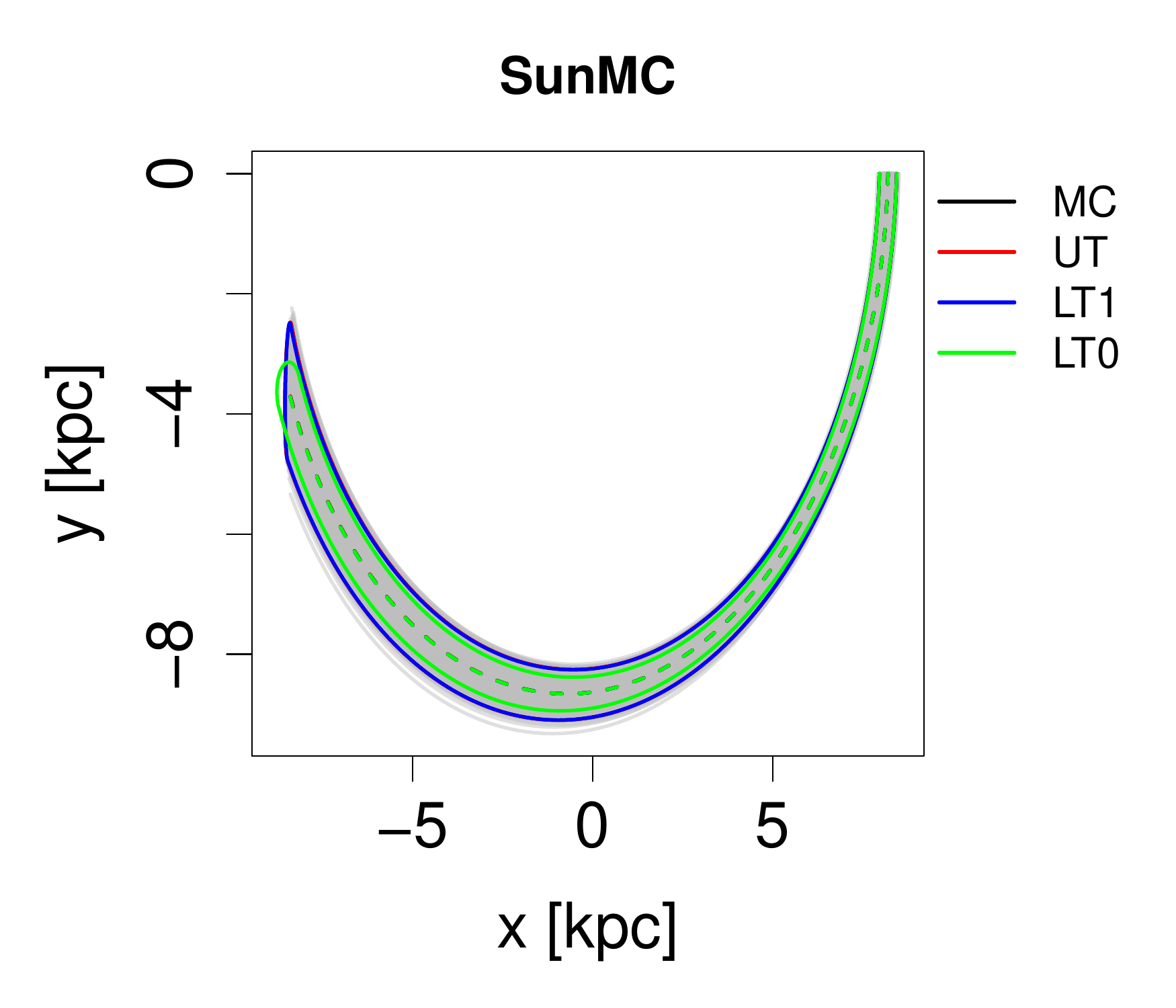}
  \includegraphics[scale=0.40]{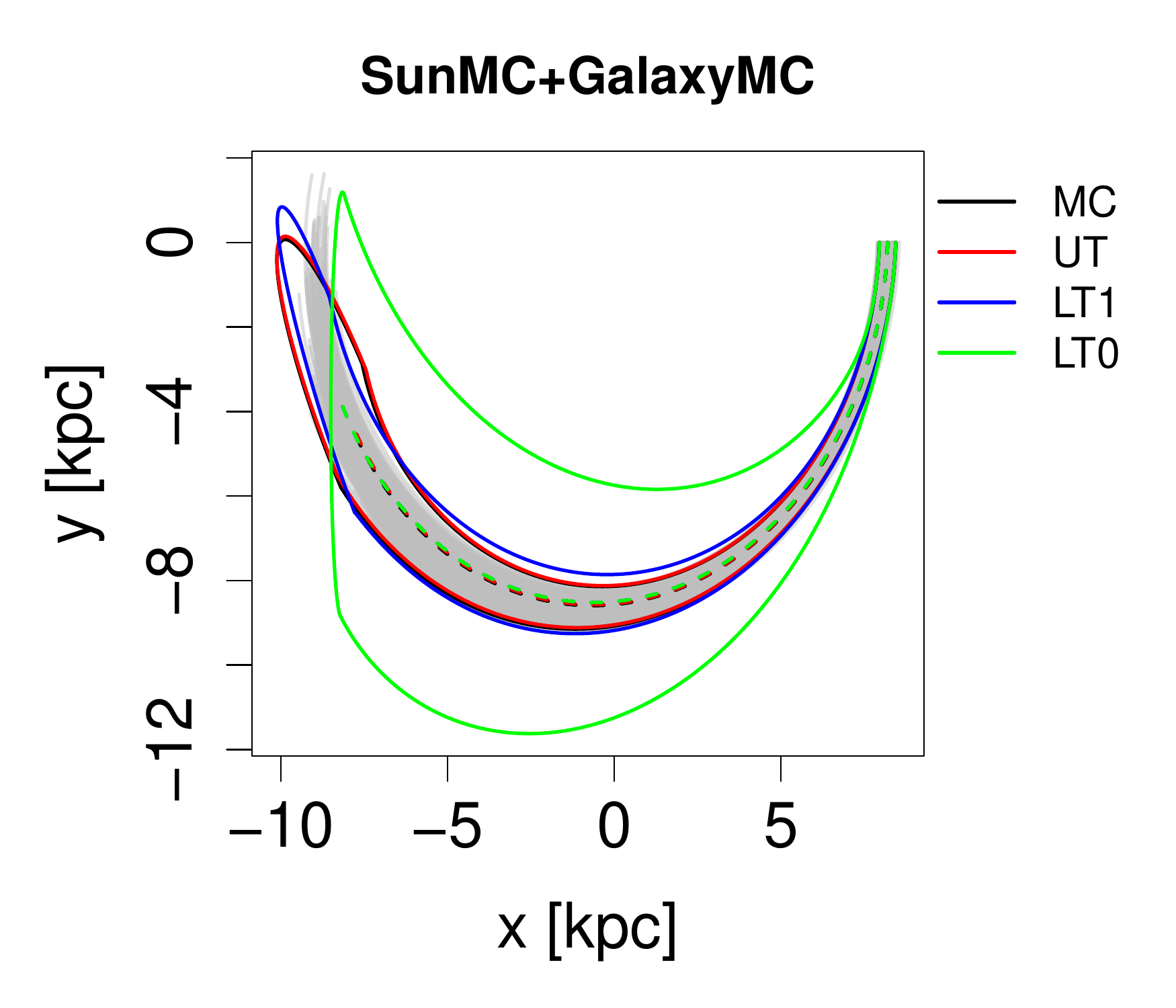}
  \caption{Comparison of the MC, UT, LT1, and LT0 methods in uncertainty propagation of the Solar orbit. The grey lines show the orbits of solar clones. The dashed lines show the mean orbit while the solid lines show the tangential point between the mean orbit and the contour of 95\% quantile. The semi-ellipse contour for the last time step is also shown. The contours for MC, UT and LT1 overlap with each other and thus are not distinguishable in the left panel. }
  \label{fig:ci}
\end{figure}

We compare the mean and covariance estimated through the MC, UT, LT1, and LT0 methods for the Solar orbit over 100\,Myr. This comparison is performed for the SunMC and SunMC+GalaxyMC experiments and is shown in Fig. \ref{fig:ci}. For the SunMC experiment, the orbital uncertainty is relatively small and thus all methods except LT0 estimate similar covariance variation. However, for the SunMC+GalaxyMC experiment, LT0 is unable to predict the covariance of the Solar orbit reliably over a few kpc scale due to the assumption of zero uncertainty of acceleration. On the other hand, the LT1 and UT methods are able to recover the covariance even over half the Sun's orbital period. The mean orbits predicted by different methods are consistent with each other. 

\begin{figure}
  \centering
  \includegraphics[scale=0.6]{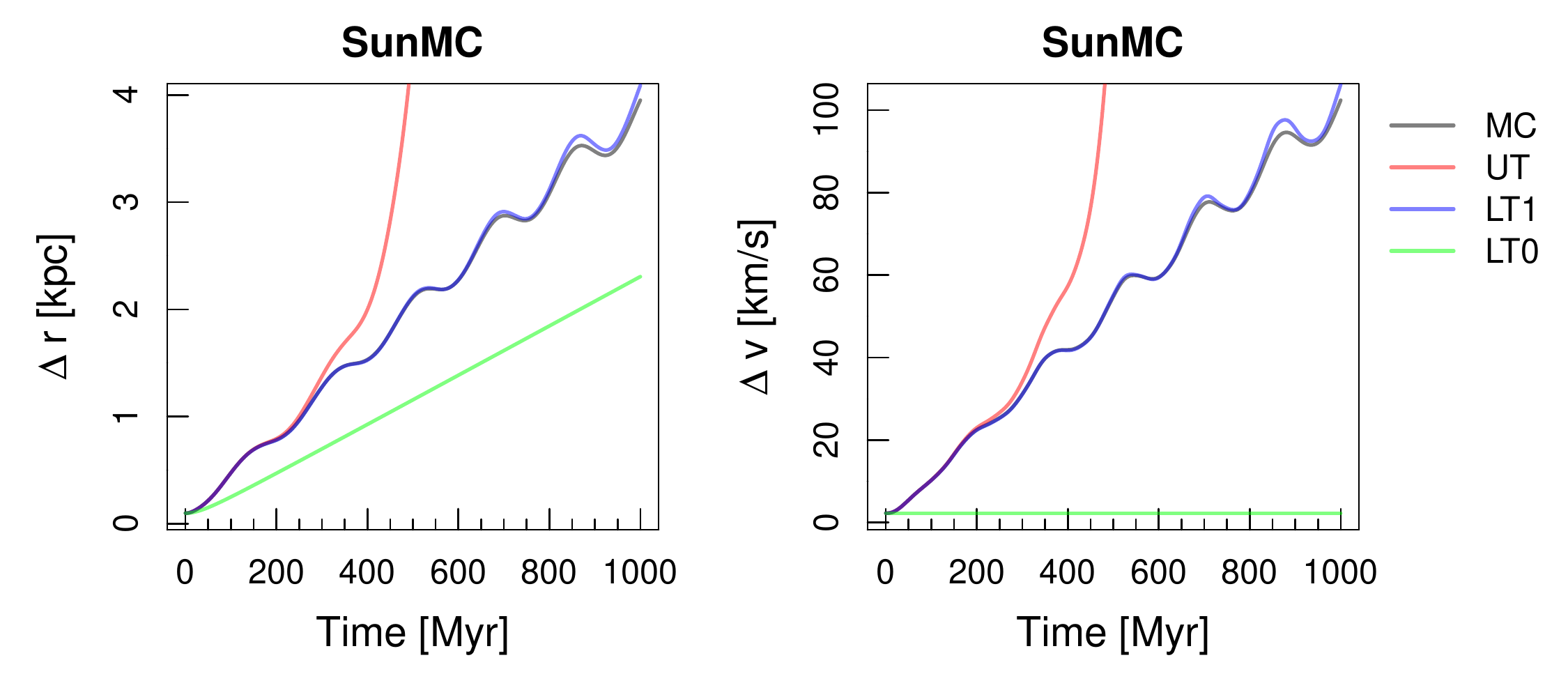}
  \includegraphics[scale=0.6]{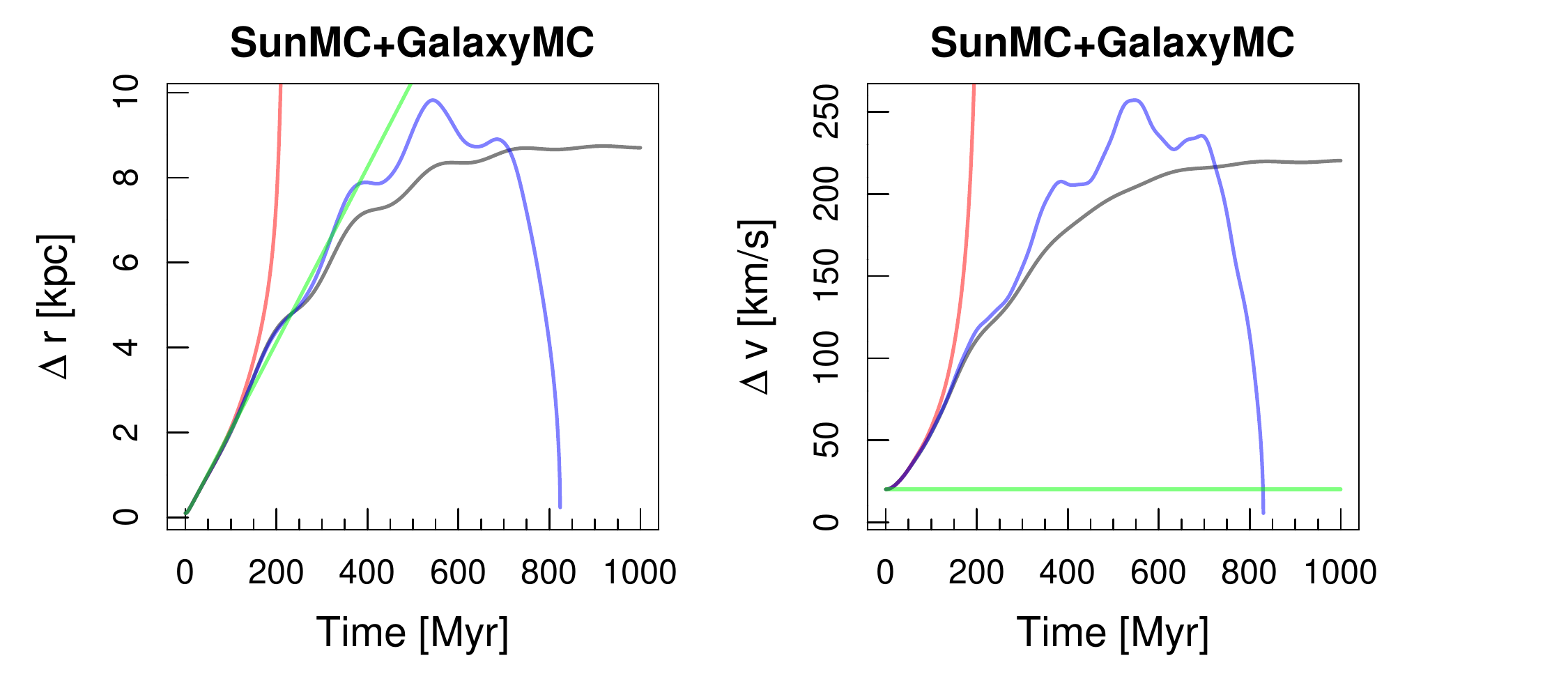}
  \caption{Comparison of the MC, UT, LT1, and LT0 estimation of the propagation of spatial (left) and velocity (right) uncertainty over 1\,Gyr for the SunMC (upper) and SunMC+GalaxyMC (lower) experiments. The spatial or velocity uncertainty is measured by the square root of the sum of the three eigenvalues of the corresponding covariance matrix. Negative and high uncertainty data are not shown to optimize visualization.}
  \label{fig:drv}
\end{figure}

To compare the performance of the different methods over a longer time, we show the propagation of the spatial and velocity uncertainty over 1\,Gyr in Fig. \ref{fig:drv}. As expected, LT0 estimates constant velocity dispersion (light green line in the right panels) due to the neglect of acceleration uncertainty. Thus a consideration of acceleration uncertainty is essential for uncertainty propagation in galactic dynamics. On the other hand, the LT1 predicts the orbital uncertainty to a relative precision of a few percent over 1\,Gyr for the SunMC experiment. The UT uncertainty diverges from the MC reference value after 200$\sim$300\,Myr integration. For the SunMC+GalaxyMC experiment, the UT is only reliable for uncertainty propagation within 100\,Myr while the LT1 is reliable for at least 200 Myr.

In contrast to previous expectations \citep{julier04,chen12}, our experiments show that the LT1 is favored over the UT in probabilistic galactic dynamics over a few hundred million year time scale for stars in the Solar neighborhood. The failure of UT over long orbital integrations is probably due to the
the cumulative error in the propagation of mean values. In the experiment of SunMC+GalaxyMC, we find the mean orbit diverges from the nominal orbit significantly after 100\,Myr. However, the LT1 only propagates the covariance but does not update the mean. A modification of the UT propagation of the mean state may improve its precision. In summary, for absolute uncertainty propagation in galactic dynamics, we recommend the LT1 for orbital uncertainty propagation over Gyr time scale for small initial uncertainty problems (such as the SunMC experiment) and for propagation over a few million years for larger initial uncertainty problems (such as the SunMC+GalaxyMC experiment). 

\subsection{Selecting the optimal non-MC method – case study of GJ 710}\label{sec:gl710}
Since the Solar orbit becomes highly unpredictable over timescales of 100 million year time scale, we consider the precision of orbital integration over shorter timescale for the stars in the Solar neighborhood. Of particular interest is GJ 710 a K dwarf about 19\,pc away from the Sun \citep{gaiaDR2}, found to pass the Sun at a distance of 0.06\,pc in about 1.3\,Myr from the present \citep{berski16,bailer-jones17,feng17pair,marcos18}. Gaia DR2 provides high precision astrometry for GJ 710, with parallax $\widetilde{\omega}=52.5185\pm0.0478$\,mas, proper motion $\mu_\alpha=-0.460\pm0.084$\,mas/yr, $\mu_\delta=-0.028\pm0.073$\,mas/yr, and radial velocity of $v_r=-14.53\pm0.44$\,km/s. Thus it provides a benchmark case for the study of orbital uncertainty over a few Myr. Although previous studies accounted for the observational uncertainty in orbital integrations, the uncertainty in the galactic model and the Sun's initial conditions increase the uncertainty considerably.

Similar to our study of the Solar orbit, we study the uncertainty of the orbit of GJ 710 using six Monte Carlo experiments, ``ObsMC'', ``SunMC'', ``GalaxyMC'', ``ObsMC+SunMC'', ``Obs+GalaxyMC'', ``ObsMC+SunMC+GalaxyMC'', where the ObsMC experiment varies observables and the other labels are the same as those previously used in section \ref{sec:sun}. For each experiment, we simulate 1000 clones of GJ 710 and the Sun over 100\,Myr. In this subsection, we first use the MC method to predict the uncertainty propagation in the Galactocentric and heliocentric reference frame for different experiments. Thus we can investigate the optimal non-MC method for the propagation of relative errors. 

\subsubsection{Absolute and relative uncertainty}\label{sec:abs_rel}
\begin{figure}
  \centering
  \includegraphics[scale=0.6]{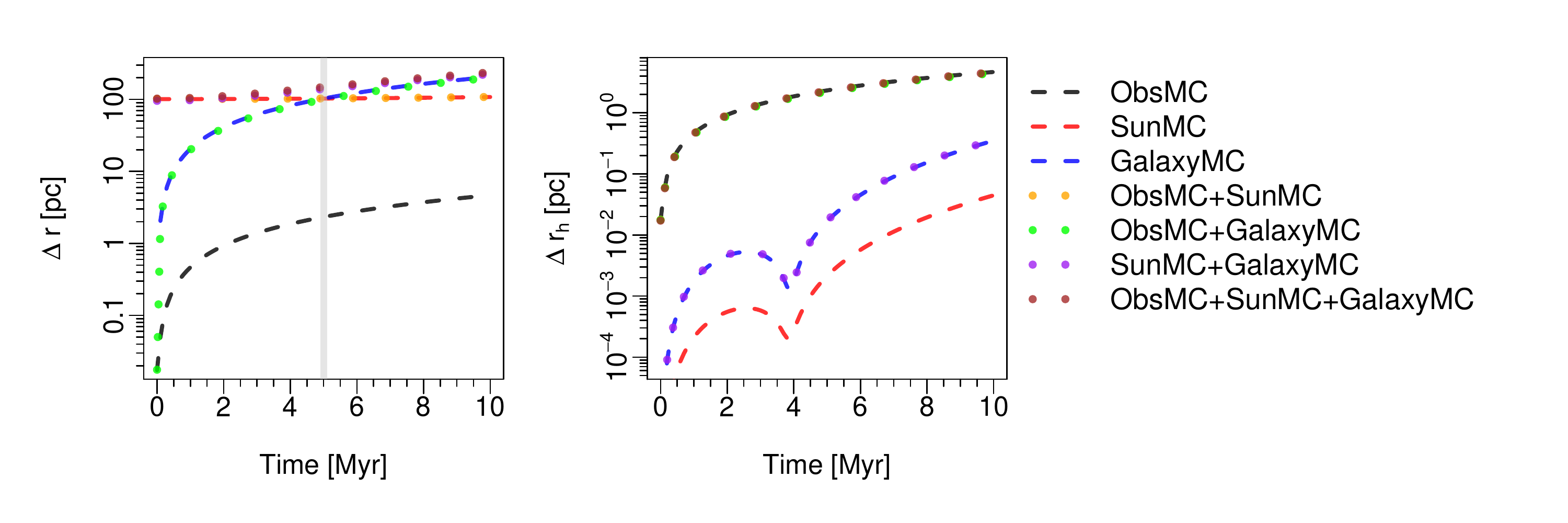}
  \includegraphics[scale=0.6]{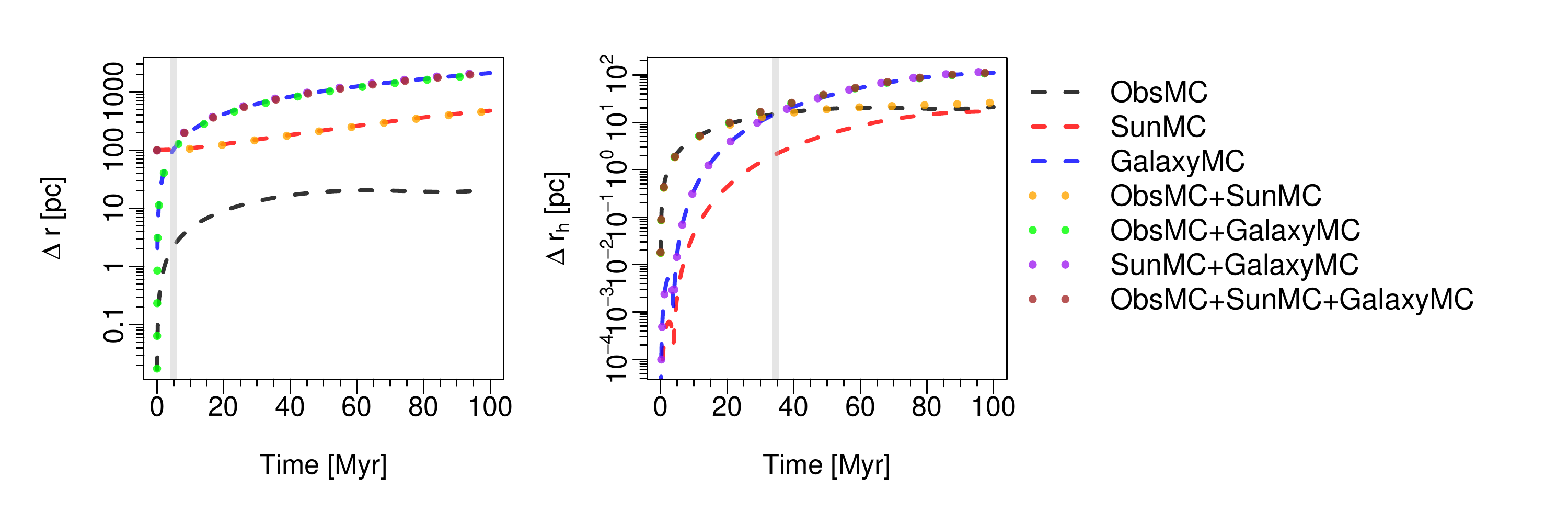}
  \caption{Spatial dispersion of the GJ 710 clones in the galactic (left panels) and heliocentric (right panels; with subscript ``h'' in the y axis label) reference frame over the past 10\,Myr (upper panels) and 100\,Myr (lower panels). The grey bars show the time when $\Delta r_h$ for GalaxyMC is equal to that for SunMC (left panels) and for ObsMC (bottom right panel). }
  \label{fig:gl710}
\end{figure}

Following the methodology discussed in Section \ref{sec:lt}, we define the uncertainty
of the orbit of GJ 710 in the galactic reference frame as ``absolute error'' and the orbital uncertainty in the heliocentric frame as ``relative error''. Based on the MC results, we show the propagation of the absolute and relative errors in Fig. \ref{fig:gl710}. These plots allow identification and quantification of the key uncertainties at different times, we do this with respect to the total relative uncertainty. This is a sum of the observable and galactic model uncertainty, as seen in the ObsMC+GalaxyMC and ObsMC+SunMC+GalaxyMC experiments. We see that the Galactocentric orbit
is very uncertain even at $t=0$\,Myr due to model uncertainty (left panels, arising from the uncertainty in the Sun's distance from the galactic centre). The initial conditions of the Sun are the main source of orbital uncertainty before 5\,Myr while the galactic model contributes the most uncertainty after 5\,Myr. However, the heliocentric orbit of GJ 710 is precise to a few parsecs within the 10\,Myr integration (upper right panel). The lower-right panel indicates that observables are the main source of relative uncertainty before 34.4\,Myr while the galactic model uncertainty contributes most relative uncertainty at later times. The initial conditions of the Sun become important around 100 Myr. 

\begin{figure}
  \centering
  \includegraphics[scale=0.6]{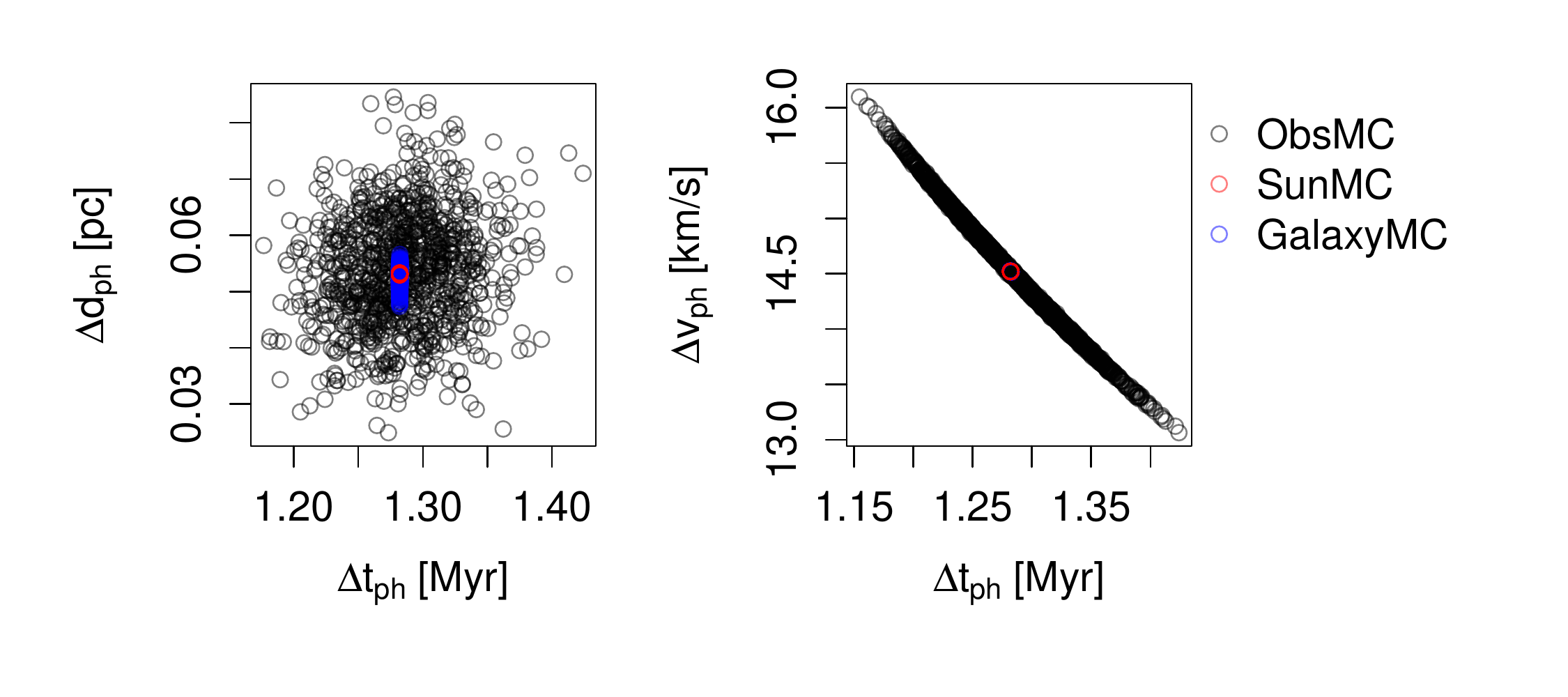}
  \caption{Distribution of the GJ 710 clones over $t_{\rm ph}$, $v_{\rm ph}$, and $r_{\rm ph}$ for the ObsMC, SunMC and GalaxyMC experiments. The clones in the SunMC and GalaxyMC experiments overlap and thus symbols are not distinguishable in the right panel. }
  \label{fig:perihelion}
\end{figure}

For each experiment, we calculate the closest distance between GJ 710 and the Sun using a linear approximation to refine the perihelion, following \cite{bailer-jones17} and \cite{feng17pair}. Since the uncertainty in the combined MC experiments is always dominated by the main MC component, we compare the perihelion time $t_{\rm ph}$, distance $d_{\rm ph}$ and velocity $v_{\rm ph}$ for the single MC experiments, ObsMC, SunMC, and GalaxyMC, in Fig. \ref{fig:perihelion}. Similar to what we find in Fig. \ref{fig:gl710}, the observables are the dominant uncertainty source for the encounter parameters. However, the galactic model also contributes considerable uncertainty to $d_{\rm ph}$, as seen from the left panel of Fig. \ref{fig:perihelion} which is scaled to focus on the expected 1.3\,Myr closest passage of GJ710 to the Sun. The standard deviation of $d_{\rm ph}$ for the ObsMC, SunMC, and GalaxyMC experiments are 9.7$\times 10^{-3}$, 5.4$\times 10^{-5}$, and 1.4$\times 10^{-3}$\,pc, respectively. The standard deviation of $d_{\rm ph}$ for ObsMC+GalaxyMC is 9.8$\times 10^{-3}$\,pc and is the quadrature-sum of the uncertainty in the ObsMC and GalaxyMC experiments. 

Hence the galactic model makes a considerable contribution to the uncertainty in the orbit and because the uncertainties in the galactic model and in the Sun’s initial conditions are independent of the observational uncertainty, the orbital uncertainty caused by them should be added in quadrature to the observation-based orbital uncertainty for uncertainty. 

\subsubsection{Methods of relative uncertainty propagation}\label{sec:relative}
\begin{figure}
  \centering
  \includegraphics[scale=0.6]{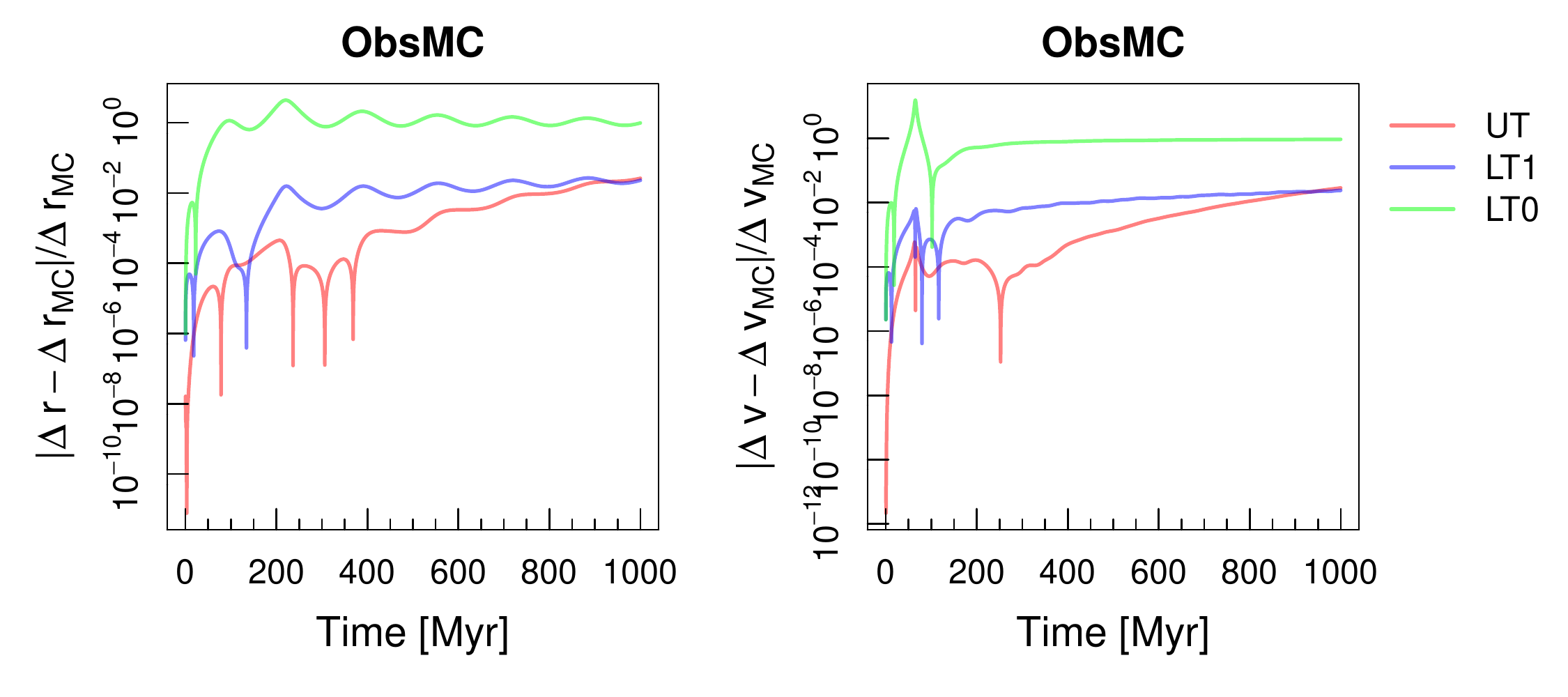}
  \includegraphics[scale=0.6]{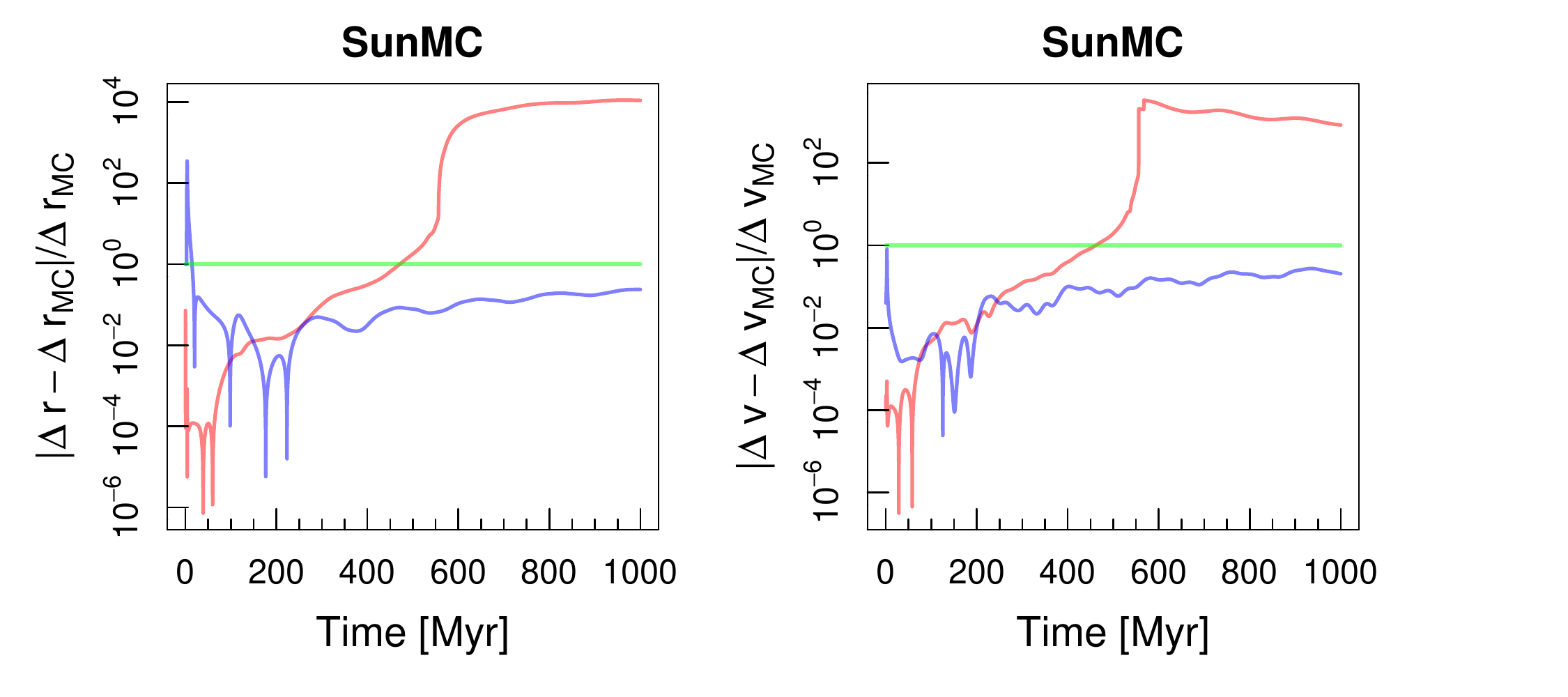}
  \includegraphics[scale=0.6]{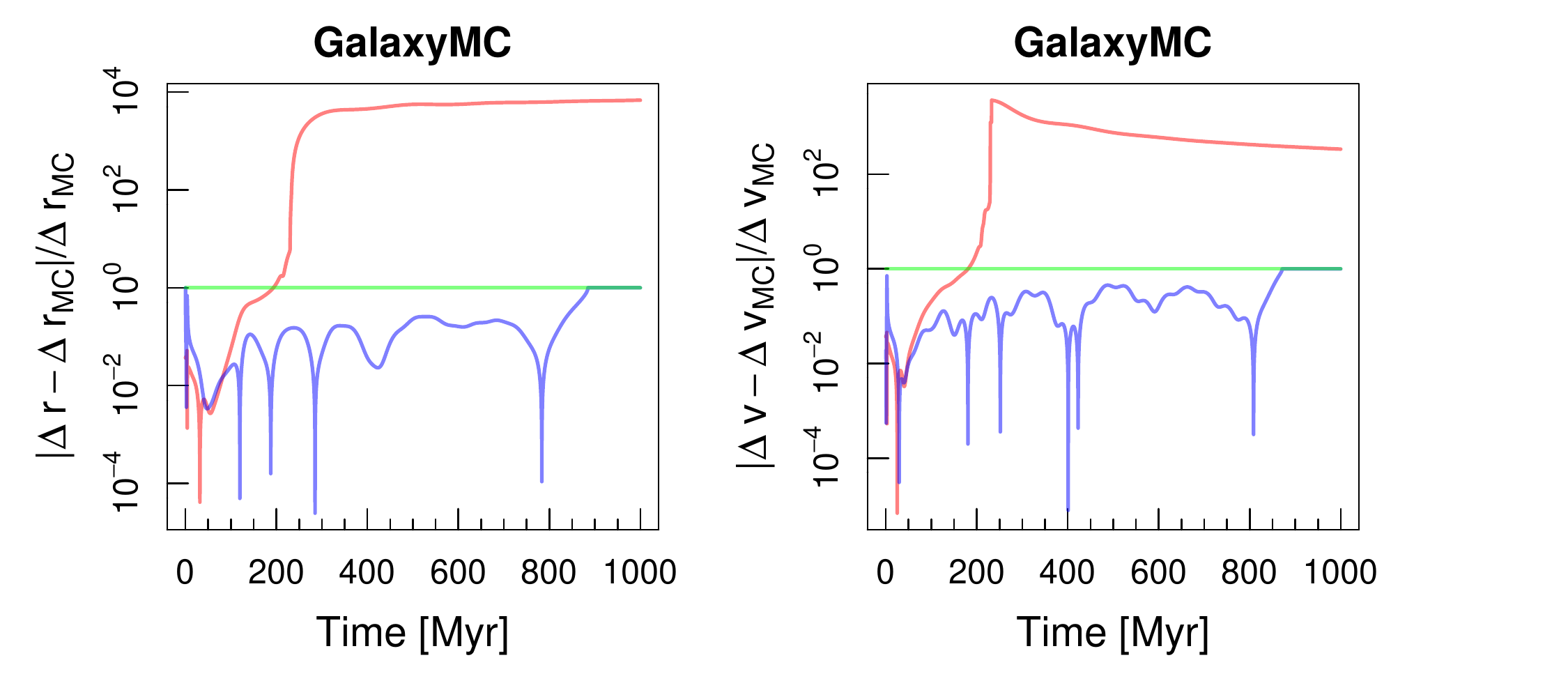}
  \caption{Relative spatial (left) and velocity (right) uncertainty propagation of GJ 710's orbit over 1\,Gyr. The orbital uncertainties are calculated using the UT, LT1, and LT0 methods and are compared with the MC-predicted uncertainty}
  \label{fig:relative}
\end{figure}

In the context of galactic dynamics, one typically aims at determining the uncertainty of one orbit relative to another. We use the GJ 710 as an example to compare the different uncertainty propagation methods. As in section \ref{sec:absolute}, we compare the UT, LT1 and LT0 methods with respect to the MC method for various MC experiments with an integration time step of 0.1\,Myr. As found in the previous subsection, the orbital uncertainty is typically dominated by a single MC component. Thus we only compare propagation methods for single MC experiments, namely ObsMC, SunMC, and GalaxyMC and show the relative spatial and velocity uncertainty in Fig. \ref{fig:relative}. In the upper panels, we see that the UT and LT1 predicts the spatial and velocity uncertainty to a precision of 1\% over 1\,Gyr. The UT even predicts the uncertainty to a 0.01\% precision over the first 400\,Myr. However, for the SunMC experiment, the UT uncertainty significantly diverges from the MC reference value after 300\,Myr (middle panels). The UT uncertainty diverges from the MC uncertainty even earlier for the GalaxyMC experiment (lower panels). On the other hand, the LT1 performs reasonably well over the whole 1\,Gyr integration time span. Since the LT0 does not account for acceleration uncertainty, it predicts constant relative uncertainty for the SunMC and GalaxyMC experiments. It also gives a poor uncertainty prediction for the ObsMC experiments. Hence the UT is the most accurate method for the propagation of relative uncertainty within one hundred million years while the LT1 is appropriate for uncertainty propagation over a longer timescale. In other words, the UT is optimal for small uncertainty propagation while the LT1 is optimal for relatively larger uncertainty propagation, considering that the orbital uncertainty is increasing over time. 

The other metric used to measure the performance of uncertainty propagation is the computation time. We show the computation times of different propagation methods for the ObsMC, SunMC and GalaxyMC experiments in Table \ref{tab:time}. The computation time of UT is about one order of magnitude longer than the LT methods. The MC takes about two orders of magnitude longer time than the UT method.
For a system with $n$ parameters, if each dimension is sampled with 10 points, the operations of MC are $\mathcal{O}(10^n)$. However, the operations for UT and LT are respectively $\mathcal{O}((2n+1)^3)$ \citep{julier04} and $\mathcal{O}(1)$. Although the MC method is the most accurate one for uncertainty propagation, it is computationally expensive and is even not feasible for high-dimensional nonlinear systems. Unlike the MC, the UT samples the sigma points deterministically. The number of sigma points is a linear function of the system dimension rather than an exponential function of the dimension as in the MC method. Overall, the LT1 is the fastest way to accurately predict orbital uncertainty in a resource limited environment despite the requirement to calculate Jacobian matrices. 
\begin{table}
  \centering
  \caption{Core time (2.5 GHz Intel Core i7 in an MacBook Pro) of relative uncertainty propagation in units of seconds for various uncertainty propagation methods for the ObsMC, SunMC, and GalaxyMC  experiments. In each MC experiment, the orbits of 1000 clones as well as the nominal orbit of GJ 710 are computed with an integration time step of 0.1\,Myr over 1\,Gyr.}
  \label{tab:time}
  \begin{tabular}{l*4{r}}
    \hline
    Experiments&MC&UT&LT1&LT0\\\hline\hline
    ObsMC&14040&139&18&18\\
    SunMC&12384&325&17&17\\
    GalaxyMC&12456&479&17&17\\\hline
    \end{tabular}
  \end{table}

  Based on the above analysis, the optimal strategy for probabilistic orbital integration should take the following factors into account.
  \begin{itemize}
  \item If there is no need to limit computational resources, the MC approach is optimal for low or moderate dimensional nonlinear uncertainty propagation. For example, the propagation of orbital uncertainty of one or two objects in the Galaxy. However, the MC method is not feasible for problems with high-dimensionality such as probabilistic integration of planetary systems or dynamics of stellar streams accounting for model uncertainties. 
  \item If computation resources are limited (e.g. less than 1000 cores and less than 2 day computation time), the UT is optimal for uncertainty propagation within 100\,Myr while the LT1 should be preferred for longer integration times. For example, the UT is preferred for a probabilistic study of stellar encounters of the Sun, which need a high precision uncertainty propagation in order to infer reliable encounter parameters. On the other hand, the LT1 is preferred for tracing the orbits of hyper-velocity stars or halo stars backward to their origin place, which typically require a few 100 Myr or less of integration time.
  \item For any initial investigation or test of a nonlinear system over short timescales, the UT is straightforward and does not require calculation of Jacobian matrices.
  \item For all cases, the LT1 can predict relatively high precision uncertainty propagation despite considerable initial effort in the calculation of the gradients of acceleration. 
    \end{itemize}

\section{Discussions and conclusions}\label{sec:conclusion}
We study the probabilistic dynamics of the Sun integrating not only its orbit but also its orbital uncertainty or covariance using various MC experiments. The Sun's orbital period is $T_{\phi}=224.2\pm22.3$\,Myr, epicycle oscillation period is  $T_{\rm R}=163.2\pm16.7$\,Myr, and vertical oscillation or double mid-plane crossing period is $87.8\pm10.6$\,Myr. We find that the solar motion becomes unpredictable after 800\,Myr. However, the predictability timescale is linearly proportional to the precision of the galactic model. We look forward to an improvement in our understanding of the solar motion through a more precise modeling of the Galaxy based on the Gaia data in the near future.  

As a computationally faster alternative to our MC experiments against which our results are benchmarked we investigate the UT and LT uncertainty propagation methods to integrate the orbits of the Sun and GJ 710 as two examples to compare the performance of these methods for absolute and relative uncertainty propagation, respectively. By comparing against various MC experiments, we find that the LT1 predicts the uncertainty propagation to a relative precision as high as a few percent at least over hundreds of Myrs. The UT is optimal for extremely high precision uncertainty propagation within 100\,Myr. The precision of the UT method could be improved by increasing the precision of the propagation of the mean state for the case of galactic dynamics. 

The orbit of GJ 710 is investigated probabilistically by MC. Based on the results of multiple MC experiments, we find that the observational uncertainty is the main source for relative orbital uncertainty over a few million years. Although GJ 710 passes the Sun in only about 1.3 Myr, the galactic model contributes a considerable amount of uncertainty to the perihelion time and distance with the uncertainty of the galactic model becoming dominant. The Sun's initial condition and the galactic model are the main sources of absolute orbital uncertainty. Although GJ 710 passes the Sun in about 1.3 Myr, the galactic model contributes a considerable amount of uncertainty to the perihelion time and distance. Thus we expect even more signifiant influence of the model uncertainty on encounter parameters for stars which pass the Sun more than a few million years from the present.

The probabilistic dynamics established in this work does not account for the time variation of the galactic potential caused by substructures such as spiral arms or space-time perturbations from gravitational waves. If the substructures are analytically modeled, the corresponding uncertainty can be accounted for by the UT and LT methods in the same way as the galactic model uncertainty. However, a noise term should be included into the probabilistic dynamical to account for the stochastic perturbations from stellar encounters and gravitational waves. This is similar to the statistical noise contained in the Kalman filter and should be implemented in a future research.

In summary, we investigate the reliability of the UT and LT1 methods for the propagation of orbital uncertainty through comparison with MC experiments. This approach can be adapted for the cases of other astrophysical problems which require intensive MC simulations. For example, an MC study of the evolution of the Oort cloud requires millions of samples to be simulated over a few Gyrs (e.g. \citealt{kaib11}). An alternative approach is to model the Oort cloud structure as a mixture of multivariate Gaussian distributions (e.g. \citealt{terejanu08}) and to propagate the covariance of each Gaussian distribution over a few Gyrs using the LT1 method. Considering the promise of the UT and LT1 methods, further development is warranted for the application of them in the simulations of N-body systems and the study of other astro-dynamical systems. 

\section*{Acknowledgements}
We acknowledge support from the UK Science and Technology Facilities Council [ST/M001008/1]. All numerical experiments have been done using the University of Hertfordshire high-performance computing facility (http://stri-cluster.herts.ac.uk/). This work has made use of data from the European Space Agency (ESA) mission {\it Gaia} (\url{https://www.cosmos.esa.int/gaia}), processed by the {\it Gaia}
Data Processing and Analysis Consortium (DPAC, \url{https://www.cosmos.esa.int/web/gaia/dpac/consortium}). Funding for the DPAC has been provided by national institutions, in particular the institutions
participating in the {\it Gaia} Multilateral Agreement.

{\it Software:} {\small R} package {\small DeSolve}\citep{desolve} and {\small pracma}(\url{https://cran.r-project.org/web/packages/pracma/pracma.pdf})  
\bibliographystyle{mn2e}
\bibliography{nm}
\end{document}